\documentclass{elsart}
\usepackage{natbib}
\usepackage{graphicx}

\newcommand {\CP} {{\it Capo Passero}}

\begin{document}

\runauthor{Giorgio Riccobene, Antonio Capone}

\begin{frontmatter}

\title{Deep seawater inherent optical properties in the Southern Ionian Sea}


 \author[INFNLNS]{G. Riccobene\corauthref{ca:fax}}
 \ead{riccobene@lns.infn.it},
 \author[INFNRM]{A. Capone},
 \author[INFNCT]{S. Aiello},
 \author[INFNBA]{M. Ambriola},
 \author[INFNRM]{F. Ameli},
 \author[INFNLNS,UniCT]{I. Amore},
 \author[INFNGE]{M. Anghinolfi},
 \author[INFNLNS]{A. Anzalone},
 \author[INFNPI]{C. Avanzini},
 \author[INFNNA]{G. Barbarino},
 \author[INFNBA]{E. Barbarito},
 \author[INFNGE]{M. Battaglieri},
 \author[INFNBA]{R. Bellotti},
 \author[INFNPI]{N. Beverini},
 \author[INFNRM]{M. Bonori},
 \author[INFNPI]{B. Bouhadef},
 \author[INFNNA,INAFNA]{M. Brescia},
 \author[INFNLNS]{G. Cacopardo},
 \author[INFNBA]{F. Cafagna},
 \author[INFNCT]{L. Caponetto},
 \author[INFNPI]{E. Castorina},
 \author[INFNBA]{A. Ceres},
 \author[INFNRM]{T. Chiarusi},
 \author[INFNBA]{M. Circella},
 \author[INFNLNS]{R. Cocimano},
 \author[INFNLNS]{R. Coniglione},
 \author[INFNLNF]{M. Cordelli},
 \author[INFNLNS]{M. Costa},
 \author[INFNGE]{S. Cuneo},
 \author[INFNLNS]{A. D'Amico},
 \author[INFNRM]{G. De Bonis},
 \author[INFNBA]{C. De Marzo \thanksref{Dead}},
 \author[INFNNA]{G. De Rosa},
 \author[INFNGE]{R. De Vita},
 \author[INFNLNS]{C. Distefano},
 \author[INFNPI]{E. Falchini},
 \author[INFNBA]{C. Fiorello},
 \author[INFNPI]{V. Flaminio},
 \author[INFNGE]{K. Fratini},
 \author[INFNBO]{A. Gabrielli},
 \author[INFNPI]{S. Galeotti},
 \author[INFNBO]{E. Gandolfi},
 \author[INFNCT]{A. Grimaldi},
 \author[INFNLNF]{R. Habel},
 \author[INFNCT,UniCT]{E. Leonora},
 \author[INFNRM]{A. Lonardo},
 \author[INFNNA]{G. Longo},
 \author[INFNCT,UniCT]{D. Lo Presti},
 \author[INFNRM]{F. Lucarelli},
 \author[INFNPI]{E. Maccioni},
 \author[INFNBO]{A. Margiotta},
 \author[INFNLNF]{A. Martini},
 \author[INFNRM]{R. Masullo},
 \author[INFNBA]{R. Megna},
 \author[INFNLNS,UniCT]{E. Migneco},
 \author[INFNBA]{M. Mongelli},
 \author[INFNPI]{M. Morganti},
 \author[UniWi]{T. Montaruli \thanksref{UniBa}},
 \author[INFNLNS]{M. Musumeci},
 \author[INFNRM]{C.A. Nicolau},
 \author[INFNLNS]{A. Orlando},
 \author[INFNGE]{M. Osipenko},
 \author[INFNNA]{G. Osteria},
 \author[INFNLNS]{R. Papaleo},
 \author[INFNLNS]{V. Pappalardo},
 \author[INFNCT,UniCT]{C. Petta},
 \author[INFNLNS]{P. Piattelli},
 \author[INFNPI]{F. Raffaelli},
 \author[INFNLNS]{G. Raia},
 \author[INFNCT]{N. Randazzo},
 \author[INFNCT]{S. Reito},
 \author[INFNGE]{G. Ricco},
 \author[INFNGE]{M. Ripani},
 \author[INFNLNS]{A. Rovelli},
 \author[INFNBA]{M. Ruppi},
 \author[INFNCT,UniCT]{G.V. Russo},
 \author[INFNNA]{S. Russo},
 \author[INFNLNS]{S. Russo \thanksref{ICRAM}},
 \author[INFNLNS]{P. Sapienza},
 \author[INFNLNS]{M. Sedita},
 \author[INFNRM]{J-P. Schuller \thanksref{CEA}},
 \author[INFNGE]{E. Shirokov},
 \author[INFNRM]{F. Simeone},
 \author[INFNCT,UniCT]{V. Sipala},
 \author[INFNBO]{M. Spurio},
 \author[INFNGE]{M. Taiuti},
 \author[INFNPI]{G. Terreni},
 \author[INFNLNF]{L. Trasatti},
 \author[INFNCT]{S. Urso},
 \author[INFNLNF]{V. Valente},
 \author[INFNRM]{P. Vicini},


\corauth[ca:fax]{Fax: +39 095 542 398}
\thanks[Dead]{Deceased}
\thanks[ICRAM]{Present address: Istituto Centrale per la Ricerca Scientifica e Tecnologica Applicata al Mare, via Casalotti 300, 00166, Roma, Italy}
\thanks[CEA]{Present address: DAPNIA/SPP Bât 141 CEN Saclay, 91191 Gif-sur-Yvette, France}
\thanks[UniBa]{On leave of absence Dipartimento Interateneo di Fisica Universit\`a di Bari, Via E. Orabona 4, 70126, Bari, Italy }

\address[INFNLNS]{Laboratori Nazionali del Sud INFN, Via S.Sofia 62, 95123, Catania, Italy}
\address[INFNLNF]{Laboratori Nazionali di Frascati INFN, Via Enrico Fermi 40, 00044, Frascati (RM), Italy}
\address[INFNCT]{INFN Sezione Catania, Via S.Sofia 64, 95123, Catania, Italy}
\address[INFNBA]{INFN Sezione Bari and Dipartimento Interateneo di Fisica Universit\`a di Bari,  Via E. Orabona 4, 70126, Bari, Italy}
\address[INFNBO]{INFN Sezione Bologna and Dipartimento di Fisica Universit\`a di Bologna, V.le Berti Pichat 6-2, 40127, Bologna, Italy}
\address[INFNGE]{INFN Sezione Genova and Dipartimento di Fisica Universit\`a di Genova, Via Dodecaneso 33, 16146, Genova, Italy}
\address[INFNNA]{INFN Sezione Napoli and Dipartimento di Scienze Fisiche Universit\`a di Napoli, Via Cintia, 80126, Napoli, Italy}
\address[INFNPI]{INFN Sezione Pisa and Dipartimento di Fisica Universit\`a di Pisa, Polo Fibonacci, Largo Bruno Pontecorvo 3, 56127, Pisa, Italy}
\address[INFNRM]{INFN Sezione Roma 1 and Dipartimento di Fisica Universit\`a di Roma "La Sapienza", P.le A. Moro 2, 00185, Roma, Italy}
\address[UniCT]{Dipartimento di Fisica and Astronomia Universit\`a di Catania, Via S.Sofia 64, 95123, Catania, Italy}
\address[INAFNA] {INAF Osservatorio Astronomico di Capodimonte, Salita Moiariello 16, 80131, Napoli, Italy}
\address[UniWi]{University of Wisconsin, Department of Physics, 53711, Madison, WI, USA}


\begin{abstract}

The NEMO (NEutrino Mediterranean Observatory) Collaboration has
been carrying out since 1998 an evaluation programme of deep sea
sites suitable for the construction of the future Mediterranean
km$^3$ \v{C}erenkov neutrino telescope. We investigated the
seawater optical and oceanographic properties of several deep sea
marine areas close to the Italian Coast. Inherent optical
properties (light absorption and attenuation coefficients) have
been measured as a function of depth using an experimental
apparatus equipped with standard oceanographic probes and the
commercial transmissometer AC9 manufactured by WETLabs. This paper
reports on the visible light absorption and attenuation
coefficients measured in deep seawater of a marine region located
in the Southern Ionian Sea, 60$\div$100 km SE of Capo Passero
(Sicily). Data show that blue light absorption coefficient is
about 0.015 m$^{-1}$ (corresponding to an absorption length of 67
m) close to the one of optically pure water and it doe not show
seasonal variation. \vspace{1pc}
\end{abstract}


\end{frontmatter}

\section{Introduction}

The construction of km$^3$-scale high energy neutrino telescopes
will complement and extend the field of high energy astrophysics
allowing the identification of the highest energy cosmic ray
sources. The search for astronomical sources of high energy cosmic
rays is possible with particles that reach un-deflected the
detectors. The observational horizon of high energy cosmic gamma
rays and nuclei from ground based detectors is limited to few tens
of Mpc by the interaction with cosmic matter and radiation: this
should imply the well known GZK cutoff
\cite{GaisserHalzenStanev1995,LearnedMannheim2000} in the energy
distribution of ultra high energy extragalactic cosmic rays. On
the contrary, the low cross section of weak interaction allows
neutrinos to reach the Earth un-deflected from the farthermost
regions of the Universe. Active Galactic Nuclei
\cite{Mannheim1995}, Galactic Supernova Remnants
\cite{Protheroe1997}, Microquasars \cite{Distefano2002} and Gamma
Ray Bursters \cite{WaxmanBahcall1997} are some of the most
promising candidate of high energy muon-neutrino sources. On the
basis of high energy neutrino fluxes, calculated using
astrophysical models, neutrino detectors with an effective area of
$\simeq$ 1 km$^2$ will be able to collect, in one year, a
statistically significant number of events from point-like
astrophysical neutrino sources.

Underwater \v{C}erenkov telescopes detect high energy neutrinos
indirectly, tracking the \v{C}erenkov light wavefront radiated, in
seawater or in ice, by charged leptons originated in Charged
Current neutrino interactions. Seawater, therefore, acts as a
neutrino target and as a \v{C}erenkov radiator. An undersea
location at a depth of more than 3000 m provides an effective
shielding for atmospheric muons background and allows the
construction of such detectors, usually referred as {\it Neutrino
Telescopes} \cite{markov1961}. Two smaller scale neutrino
detectors, AMANDA and BAIKAL, have already collected and reported
candidate neutrino events \cite{AMANDAPRL2003, BAIKALICRC2003}.
AMANDA is located in the South Pole icecap \cite{AMWWW} at a depth
between 1400 and 2400 m. The present size is relatively small,
about 25000 m$^2$ for TeV muons, compared to IceCube \cite{IQWWW},
the future $km^3$ detector now under construction. BAIKAL NT-200,
the pioneer underwater detector, is deployed in the Siberian Lake
Baikal at about 1000 m depth and has a detection area close to
10$^4$ m$^2$ for TeV muons \cite{BAWWW}.

In the Northern Hemisphere, the Mediterranean Sea offers several
areas with depths greater than 3000 m; few are close to scientific
and logistic infrastructures and offer optimal conditions to
install an underwater km$^3$ neutrino telescope. The future
IceCube and the Mediterranean km$^3$ will complement each other
providing a global 4$\pi$ observation of the sky. The long light
absorption length of the Antarctic ice is expected to allow good
energy resolution, the long light effective scattering length of
the Mediterranean seawater should also allow excellent angular
resolution. Three collaborations, NESTOR \cite{NEWWW}, ANTARES
\cite{ANWWW} and NEMO \cite{NMWWW}, are presently active in the
Mediterranean Sea. NESTOR proposes the installation of a
\v{C}erenkov detector, with a tower-shaped geometry, moored a few
nautical miles off the south-west tip of the Peloponnese (Greece),
at about 4000 m depth. ANTARES is building a detector in the
vicinity of Toulon (France) at $\simeq$2450 m depth to possibly
detect astrophysical neutrinos and to demonstrate the feasibility
of a km$^3$-scale underwater neutrino telescope.

\begin{figure}[h]
\vspace*{-2.0mm}
\centerline{\includegraphics[width=10cm]{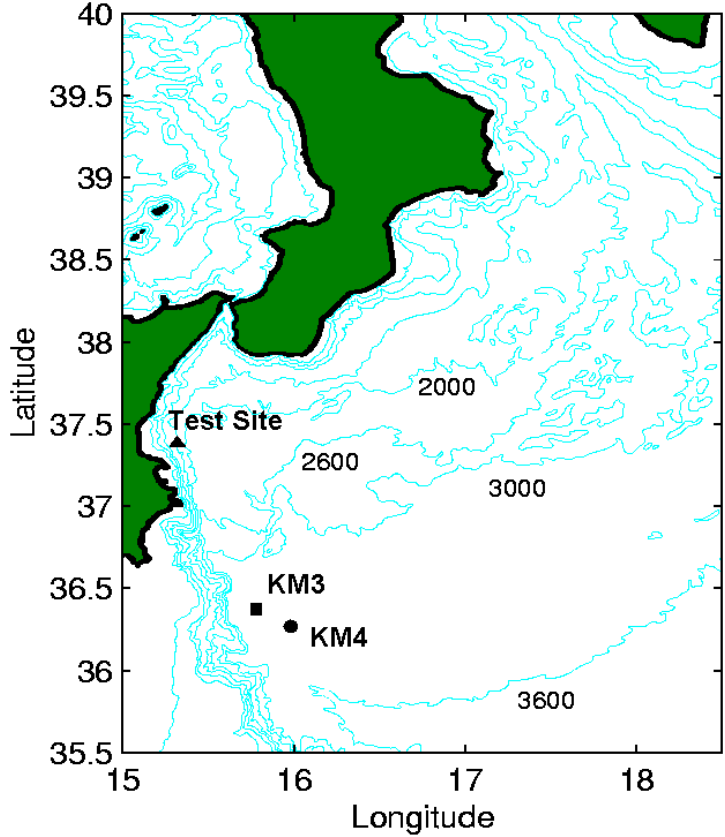}} \caption{
Bathymetric chart of the \CP~ region. The location of the $KM3$
(square) and $KM4$ (circle) sites and of the NEMO Phase 1
$Test~Site$ (triangle) is shown. The seabed depth is about 3400 m
for the \CP~ sites and 2000 m for the $Test~Site$.}
\label{fig:CapoPasseroMap}
\end{figure}

The NEMO Collaboration is active in the design and tests for the
Mediterranean km$^3$ neutrino telescope. After a long period of
$R\&D$ activity, at present the collaboration is ready to install
a prototype station (NEMO {\it phase 1}) at 2000 m depth, 25 km
offshore the town of Catania, in Sicily ({\it Test Site} in Figure
\ref{fig:CapoPasseroMap}). Since 1998 we have performed more than
25 oceanographic campaigns in the Central Mediterranean Sea in
order to characterize and eventually seek an optimal submarine
site for the installation of the Mediterranean km$^3$
\cite{NEMOHamburg2001}. Three areas close to the Italian Coast
have been compared, on the basis of two requirements: depth $>$
3000 m and distance from shore $<$ 100 km. Two of these sites are
trenches located in the Southern Tyrrhenian Sea close to the
Alicudi and Ustica Islands (at depth $\simeq$ 3500 m).
Measurements of deep seawater optical properties were performed by
the NEMO Collaboration in these sites and results were published
\cite{Capone2001}. The third site is a submarine plateau, whose
average depth is $\simeq$3500 m, located at a distance of
40$\div$100 km South East of Capo Passero, Sicily (see Figure
\ref{fig:CapoPasseroMap}). In this paper we report on deep
seawater optical properties (absorption and attenuation
coefficients) measured in the \CP~ marine region during a period
extending from December 1999 to July 2003. The results refer to
two sites located $\simeq$60 km ($36^{\circ}30$'{\bf
N},$15^{\circ}50$'{\bf E}) and $\simeq$80 km ($36^{\circ}25$'{\bf
N},$16^{\circ}00$'{\bf E}) offshore \CP, hereafter indicated
respectively as $KM3$ and $KM4$. The programme of characterisation
of deep seawater in \CP~ site, carried out by the NEMO
Collaboration, includes also long term measurements of optical
background (due to bioluminescence and $^{40}K$ radioactive
decays), water temperature and salinity, deep sea currents,
sedimentation rate and bio-fouling. The results of this work are
presented elsewhere \cite{NEMOAppec2003} and will be published
soon.

\section{Measurements of water optical properties with the AC9}

The propagation of light in water is quantified, for a given
wavelength $\lambda$, by the water Inherent Optical Properties
(IOP): the absorption $a(\lambda)$, scattering $b(\lambda)$ and
attenuation $c(\lambda)= a(\lambda) + b(\lambda) $ coefficients.
The light propagation in water can be described by the laws:

\begin{eqnarray}
\label{eq:lambertlaw}
\nonumber I_{a}(x,\lambda) = I_0 (\lambda)  e^ {-x \cdot a(\lambda)}  \\
\nonumber I_{b}(x,\lambda) = I_0 (\lambda)  e^ {-x \cdot b(\lambda)}\\
\nonumber I_{c}(x,\lambda) = I_0 (\lambda)  e^ {-x \cdot c(\lambda)}
\end{eqnarray}

\noindent where $x$ is the optical path traversed by the light and
$I_0$($\lambda$) is the source intensity. A complete description
of light scattering in water would require the knowledge of
another IOP, i.e. the scattering angular distribution, or volume
scattering function, $\tilde{\beta}$($\vartheta,\lambda$).
Integrating this function over the diffusion angle $\vartheta$ one
gets $b(\lambda)$. In this paper we shall report on measurements
of $c(\lambda)$ and $a(\lambda)$ for visible light wavelengths
performed with a commercial transmissometer, the AC9 manufactured
by WETLabs \cite{WETLabs}. It is worth to mention that the AC9
performs measurements of the attenuation coefficient in a
collimated geometry: the angular acceptance of the $c(\lambda)$
channel is $\simeq 0.7^{\circ}$. The reported values of
$c(\lambda)$ are not directly comparable with the results often
reported by other authors that concern the effective light
attenuation length (or light transmission length). This quantity
is defined as $c_{eff}(\lambda)= a(\lambda) + ( 1-\langle \cos
(\vartheta)\rangle) \cdot b(\lambda)$, where $\langle \cos
(\vartheta) \rangle$ is the average cosine of the volume
scattering function \cite{Mobley1994}.

Water IOPs are wavelength dependent: the light transmission is
extremely favoured in the range 350$\div$550 nm \cite{Mobley1994}
where the photomultipliers used in neutrino telescopes to detect
\v{C}erenkov radiation reach the highest quantum efficiency. In
natural seawater, IOPs are also function of water temperature,
salinity and dissolved particulate \cite{Pope1997,Kou1993}. The
nature of particulate, either organic or inorganic, its dimension
and concentration affect light propagation. All these
environmental parameters may vary significantly, for each marine
site, as a function of depth and time. It is important, therefore,
to perform a long term programme of {\it in situ} measurements
spanning over a long time interval \cite{Duntley1963}. It is
known, indeed, that seasonal effects like the increase of surface
biological activity (typically during spring) or the precipitation
of sediments transported by flooding rivers, enlarge the amount of
dissolved and suspended particulate, worsening the water
transparency.

\begin{figure}[htb]
\centerline{\includegraphics[width=9cm]{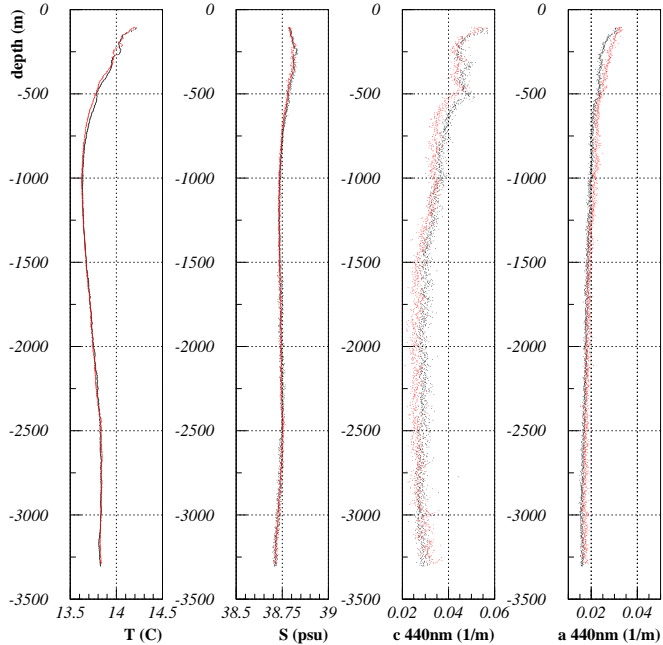}}
\caption{Temperature [$^{\circ}$C], salinity [psu], attenuation
and absorption coefficients [m$^{-1}$] for $\lambda=440$ nm as a
function of depth, measured during two deployments (red and black
dots) of the AC9 in $KM4$ site in December 1999. Results of the
measurements are nearly superimposed.} \label{fig:profileKM4}
\end{figure}

We carried out light attenuation and absorption measurements in
deep seawater using an experimental set-up based on the AC9. This
device performs attenuation and absorption measurements,
independently, using two different light paths and spanning the
light spectrum over nine different wavelengths (412, 440, 488,
510, 532, 555, 650, 676, 715 nm)
\cite{Twardowski1999,Zaneveld1994,Pegau1997}. The setup designed
for deep seawater measurements consists of an AC9, powered by a
submersible battery pack, connected to an Idronaut Ocean MK317 CTD
(Conductivity, Temperature, Depth) probe. The whole apparatus is
mounted inside an AISI-316 stainless-steel cage and it is operated
from sea surface down to deep sea, using an electro-mechanical
cable mounted on a winch onboard oceanographic research vessels.
The same cable is used to transmit the data stream to the ship
deck. The DAQ is designed to acquire, about six times per second,
water temperature, salinity, $a$($\lambda$) and $c$($\lambda$)
($412<\lambda< 715$ nm). The apparatus is typically deployed at
$\sim 0.7$ m/s vertical speed, allowing the acquisition of roughly
10 data samples per metre of depth \cite{Capone2001,Balkanov2002}.
As an example we show in Figure \ref{fig:profileKM4} the profiles,
as a function of depth, of salinity (in practical salinity units
[psu]), temperature ([$^{\circ}$C]), $a$($\lambda$=440 nm) and
$c$($\lambda$=440 nm) ([m$^{-1}$]) measured in two deployments at
the $KM4$ site during December 1999. Each plotted point represents
the average value over 10 m depth. The two measurements (red dots
and black dots), carried out in two consecutive days, are nearly
superimposed. The Figure indicates that deep waters in $KM4$ do
not show relevant variations of oceanographic and optical
properties in the depth interval 2000$\div$3250 m.

\subsection{AC9 Calibration and systematic errors}

As described in previous papers \cite{Capone2001, Balkanov2002},
the AC9 measures the difference between the absorption and
attenuation coefficients of seawater with respect to the
coefficients for pure water. The AC9 manufacturer provides a set
of instrument calibration coefficients, that refer to the
instrument response to pure water and dry air, used to obtain the
absolute values of $a$($\lambda$) and $c$($\lambda$). In order to
reduce systematic uncertainties associated to the measurements,
during each naval campaign, the AC9 calibration coefficients have
been verified several times (before and after each deployment),
recording the instrument readings for light transmission in high
purity grade nitrogen atmosphere. With this calibration procedure
we estimated that systematic errors amount to $\simeq 1.5 \times
10^{-3}$ [m$^{-1}$] for the  $a(\lambda)$ and $c(\lambda)$
measurements. We performed in each site at least two deployments
of the AC9 setup at short time interval (typically less than 1
day).

\section{Comparison of deep sea sites in the Central Mediterranean Sea}

\begin{figure}[hb]
\centerline {\includegraphics[width=9 cm]{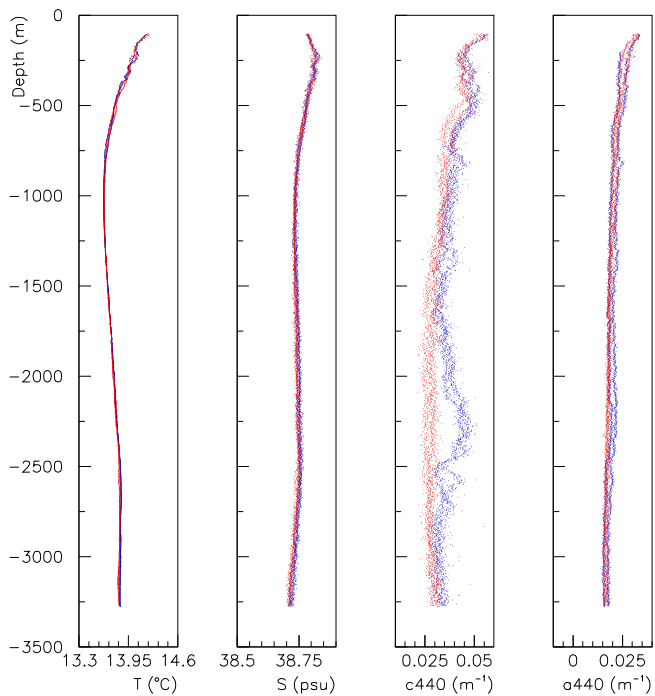}}
\caption{Comparison between temperature, salinity, attenuation and
absorption coefficients (at $\lambda=440$ nm) as a function of
depth, measured in $KM4$ (red dots) and $KM3$ (blue dots) during
December 1999. Two deployments were carried out in each site,
typically in a time window of 24 hours.}
 \label{fig:ProfileKM3_KM4}
\end{figure}

The first measurements of IOP in \CP~ were carried out in December
1999, in the $KM3$ and $KM4$ sites. A comparison among the
vertical profiles of salinity, temperature, $a$(440 nm) and
$c$(440 nm) as a function of depth, recorded in the two sites is
shown in Figure \ref{fig:ProfileKM3_KM4}. Between 1250 m and 3250
m depth, the water column in the site $KM3$ shows variations of
the attenuation coefficients as a function of depth. We attribute
this variation of $c$($\lambda$) to extra sources of light
scattering, due to particulate present in this site, which is
close to the Maltese shelf break. We never observed this effect in
$KM4$, a site farther from the Maltese Escarpment. Figure
\ref{fig:ProfileKM3_KM4}, indeed, shows that optical properties
measured in $KM4$ are almost constant as a function of depth (for
depth $>1500$ m). Table \ref{tab:Dec99_ac} summarises the values
of $a(\lambda)$ and $c(\lambda)$, measured at the $KM3$ and $KM4$
sites, averaged over an interval of about 400 m depth, 150 m above
the seabed ($\simeq$3400 m in $KM4$), which is a suitable range
for the installation of neutrino telescopes. As explained above
two deployments were carried out in each site. Results are
reported in the table. During deployments about 10 data
acquisitions per metre of depth are recorded, this implies that
large statistics is collected with the instrument in a 400 m depth
interval allowing to achieve small statistical errors for each
absorption or attenuation coefficient. Table \ref{tab:Dec99_ac}
does not report the measured values of the $a$(676 nm) coefficient
since its value is used in the off-line analysis as a
normalization parameter to estimate corrections due to the not
perfect reflectivity mirror in the AC9 absorption channel (see
reference \cite{Capone2001}). During the December 1999 sea
campaign, the attenuation channel at $\lambda$=555 nm was not
properly working therefore the $c$(555 nm) value is not given in
Table \ref{tab:Dec99_ac}.

\begin{small}
\begin{center}
\begin{table}[h]
\caption{December 1999 data. Average values of $a(\lambda)$ and
$c(\lambda)$ (in units of [m$^{-1}$]) measured in the $KM3$ and
$KM4$ sites, in the depths interval 2850$\div$3250 m. The
statistical errors are the RMS of the measured distributions. Two
deployments were carried out in each site. The systematic errors
associated with the absorption coefficient data, in all the
following tables, are of the order of $1.5 \times
10^{-3}[m^{-1}]$.} \label{tab:Dec99_ac}
\begin{tabular}{cccccc}
\hline
  coefficient & KM3                  & KM3                  & KM4                  & KM4 \\
              & $1^{st}$ measurement & $2^{nd}$ measurement & $1^{st}$ measurement & $2^{nd}$ measurement\\
\hline
  a412 & $0.0168\pm0.0006$ & $0.0137\pm0.0004$ & $0.0143\pm0.0006$ & $0.0149\pm0.0008$ \\
  a440 & $0.0177\pm0.0005$ & $0.0156\pm0.0005$ & $0.0159\pm0.0005$ & $0.0172\pm0.0007$ \\
  a488 & $0.0217\pm0.0004$ & $0.0209\pm0.0004$ & $0.0208\pm0.0004$ & $0.0213\pm0.0005$ \\
  a510 & $0.0370\pm0.0004$ & $0.0365\pm0.0004$ & $0.0363\pm0.0003$ & $0.0374\pm0.0005$ \\
  a532 & $0.0532\pm0.0004$ & $0.0527\pm0.0004$ & $0.0528\pm0.0003$ & $0.0529\pm0.0005$ \\
  a555 & $0.0682\pm0.0005$ & $0.0683\pm0.0005$ & $0.0683\pm0.0004$ & $0.0689\pm0.0006$ \\
  a650 & $0.3557\pm0.0003$ & $0.3560\pm0.0003$ & $0.3564\pm0.0003$ & $0.3581\pm0.0003$ \\
  a715 & $1.0161\pm0.0003$ & $1.0165\pm0.0003$ & $1.0167\pm0.0003$ & $1.0169\pm0.0003$ \\
\hline
  c412 & $0.0359\pm0.0025$ & $0.0336\pm0.0022$ & $0.0309\pm0.0017$ & $0.0343\pm0.0026$ \\
  c440 & $0.0335\pm0.0024$ & $0.0312\pm0.0022$ & $0.0284\pm0.0016$ & $0.0292\pm0.0025$ \\
  c488 & $0.0368\pm0.0024$ & $0.0341\pm0.0021$ & $0.0309\pm0.0015$ & $0.0329\pm0.0023$ \\
  c510 & $0.0442\pm0.0024$ & $0.0417\pm0.0020$ & $0.0397\pm0.0014$ & $0.0427\pm0.0021$ \\
  c532 & $0.0546\pm0.0024$ & $0.0520\pm0.0020$ & $0.0489\pm0.0014$ & $0.0514\pm0.0020$ \\
  c650 & $0.3780\pm0.0024$ & $0.3740\pm0.0020$ & $0.3719\pm0.0016$ & $0.3747\pm0.0022$ \\
  c676 & $0.4494\pm0.0021$ & $0.4508\pm0.0018$ & $0.4489\pm0.0011$ & $0.4503\pm0.0018$ \\
  c715 & $1.0209\pm0.0020$ & $1.0193\pm0.0018$ & $1.0169\pm0.0012$ & $1.0190\pm0.0018$ \\
\hline
\end{tabular}
\end{table}
\end{center}
\end{small}

\begin{figure}[h]
\centerline
{\includegraphics[width=7.5cm]{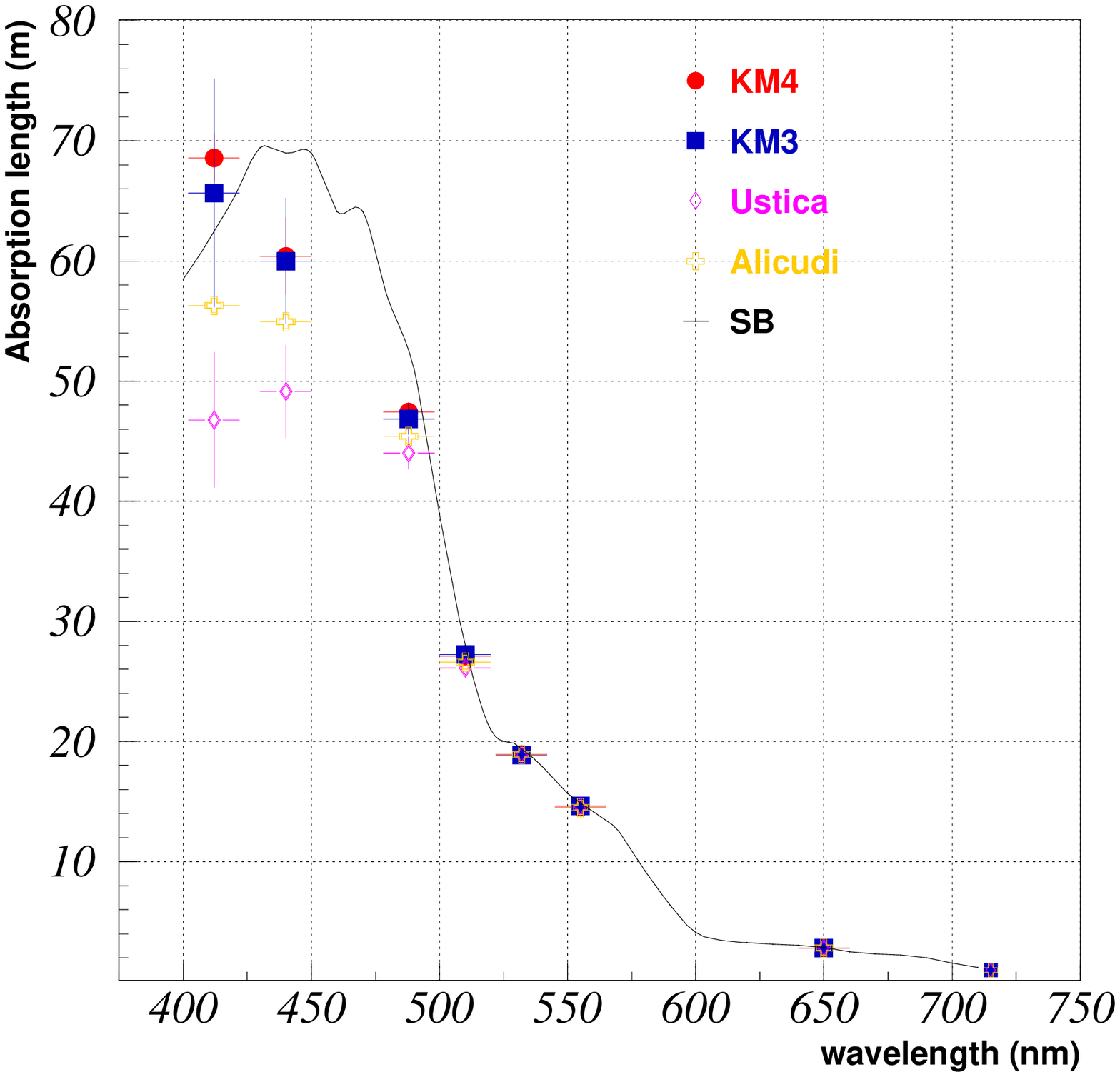}\includegraphics[width=7.5cm]{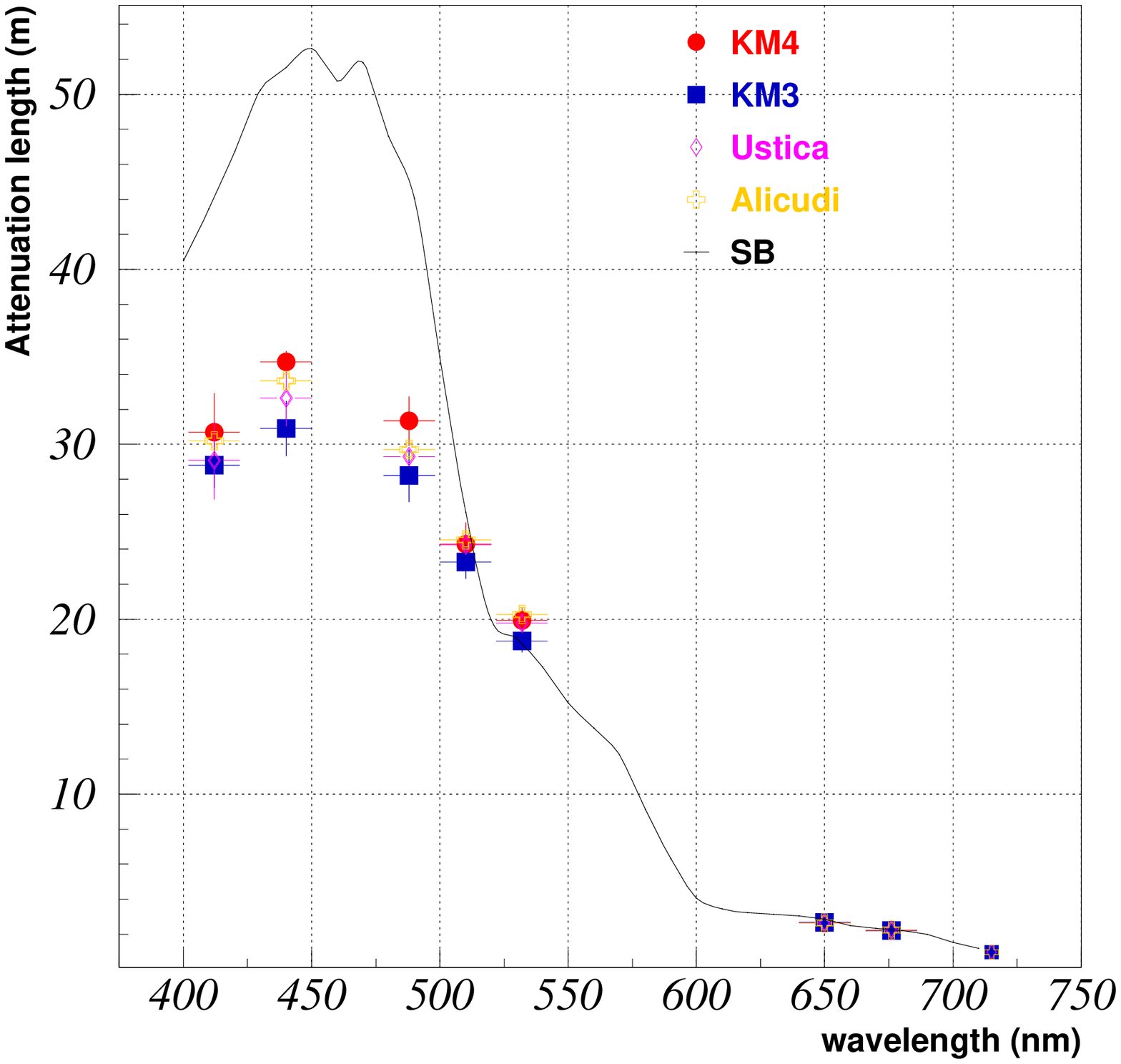}}
\caption{Average absorption and attenuation lengths measured with
the {\it AC9} in $Ustica$, $Alicudi$ (\cite{Capone2001}), \CP~
$KM3$ and $KM4$ sites, in the 2850$\div$3250 m depth interval.
Statistical errors are plotted. $L_a(\lambda)$ and $L_c(\lambda)$
of optically pure seawater, reported by Smith and Baker
\cite{SmithBaker1981}, are indicated by a solid black line.}
\label{fig:Dec99_Lac}
\end{figure}

Figure \ref{fig:Dec99_Lac} shows the absorption and attenuation
lengths ($L_a(\lambda)=1/a(\lambda)$,
$L_c(\lambda)=1/c(\lambda)$), as a function of the wavelengths
(measured in the depth range 2850$\div$3250 m) in $Ustica$ and
$Alicudi$ (see reference \cite{Capone2001}) and at the $KM3$ and
$KM4$ sites. Data presented for each site are the averages over
two deployments; the errors are the RMS of the observed
distributions. The same Figure also shows that the values of
$L_a(\lambda)$ and $L_c(\lambda)$ measured in the region of \CP~
are larger than the ones measured in the other sites. In
particular the values of $L_a(\lambda)$ are comparable to the ones
of optically pure seawater quoted by Smith and Baker
\cite{SmithBaker1981}. These results lead us to the conclusion
that in \CP~$KM4$ site the deep seawater optical properties are
close to optically pure water ones. Absorption and attenuation
coefficients are almost constant for a large interval of depths
making this site optimal for the installation of an underwater
neutrino telescope. $KM3$ site was not considered a valid choice,
in spite of the advantage to be closer to the coast, since the
measured water optical properties are not constant along the
vertical water column: this effect is supposed to be due to the
proximity to the shelf break.

\section{Long term study of optical properties at the Capo Passero site}

\begin{figure}[h]
\centerline
{\includegraphics[width=9cm]{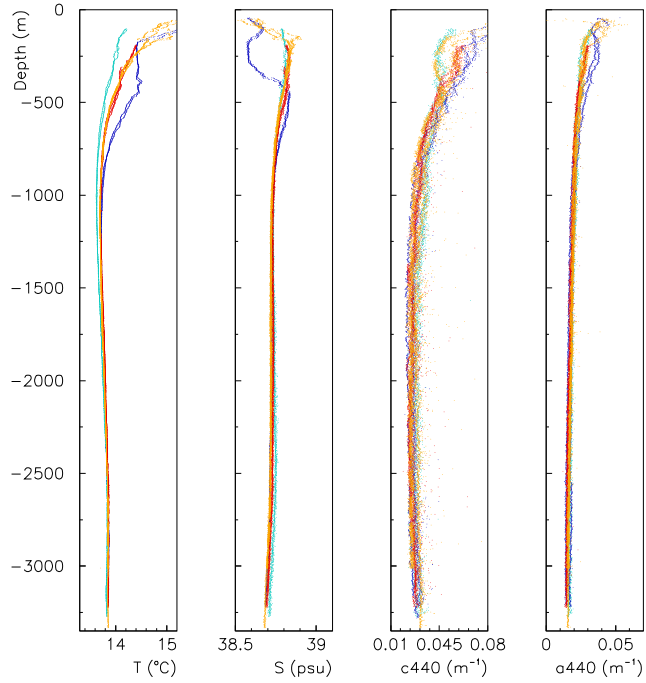}}
\caption{Profiles of temperature ($T$), salinity ($S$),
attenuation coefficient $c$(440 nm) and absorption coefficient
$a$(440 nm) measured in the \CP~ $KM4$ site. The profiles refer to
the campaigns performed during  December 1999 (2 deployments, red
dots), March 2002 (4 deployments, yellow dots), May 2002 (2
deployments, blue dots), August 2002 (3 deployments, orange dots)
and July 2003 (2 deployments, light blue dots).}
\label{fig:seasonKM4profiles}
\end{figure}

In order to verify the occurrence of seasonal variations of deep
seawater IOPs in $KM4$, we are continuously monitoring this site
using the experimental setup described above. The data collected
during oceanographic campaigns of December 1999, March 2002, May
2002, August 2002 and July 2003 are reported here. In Figure
\ref{fig:seasonKM4profiles} the profiles of water temperature,
salinity, $a$(440 nm) and $c$(440 nm), as a function of depth, are
shown. The whole collected data sample consists of: 2 deployments
in December 1999 (red dots), 4 deployments in March 2002 (yellow
dots), 2 deployments in May 2002 (blue dots), 3 deployments in
August 2002 (orange dots), 2 deployments in July 2003 (light blue
dots). Seasonal variations are observed only in shallow waters,
down to the thermocline depth of about 500 m. At depths greater
than 2000 m the $a$(440) and $c$(440) coefficients measured in
different seasons are compatible within the instrument
experimental error ($\Delta T \simeq 10^{-2}~^{\circ}$C, $\Delta S
\simeq 10^{-2}$ psu, $\Delta a,\Delta c \simeq 2.2\cdot 10^{-3}$
m$^{-1}$).

\begin{figure}[h]
\centerline
{\includegraphics[width=7.5cm]{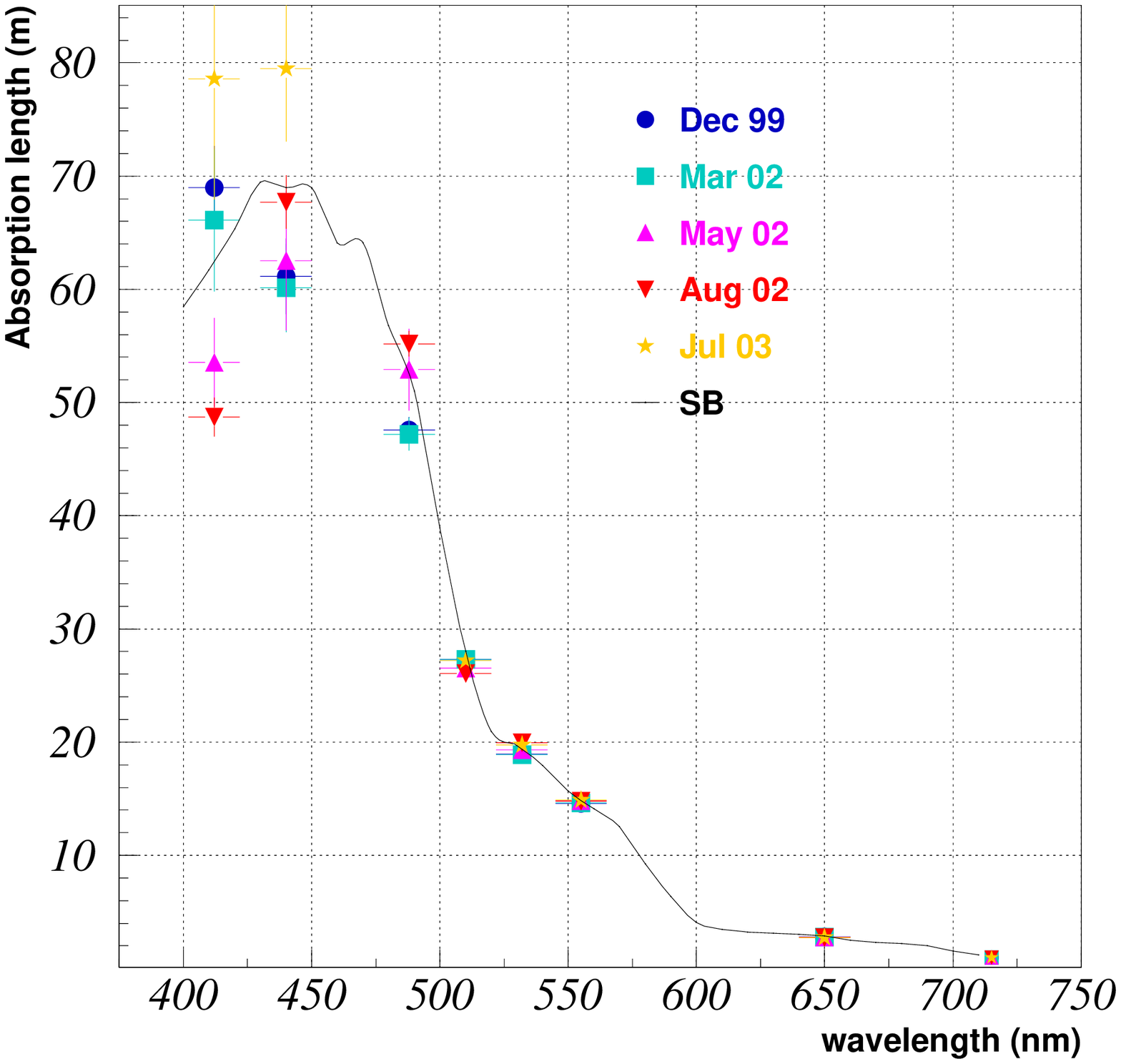}\includegraphics[width=7.5cm]{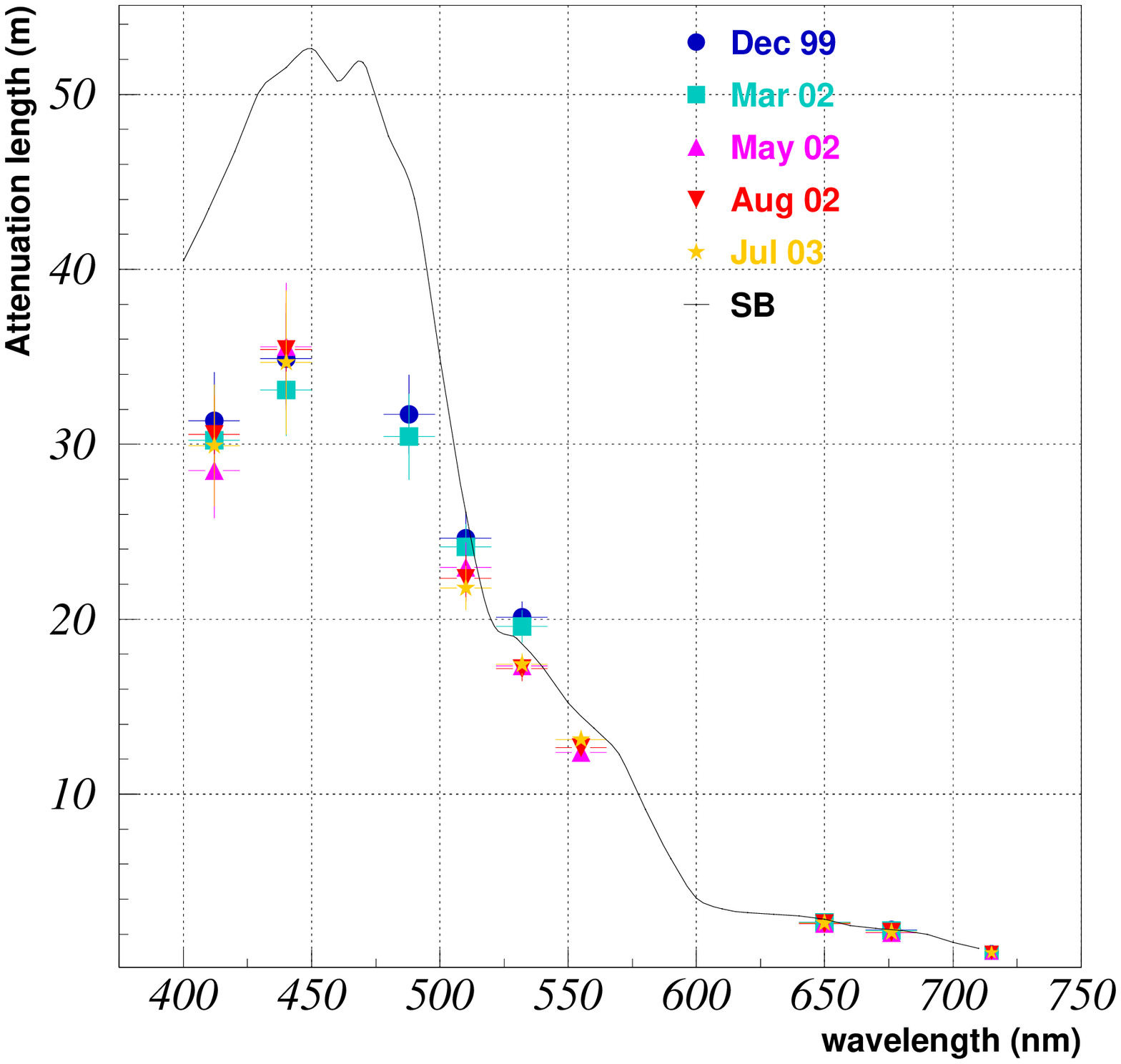}}
\caption{Average absorption and attenuation lengths measured with
the {\it AC9} in $KM4$, at depth 2850$\div$3250 m in December 1999
(blue circle), March 2002 (light blue square), May 2002 (purple
triangle), August 2002 (red upsidedown triangle) and July 2003
(dark yellow star). Statistical errors are plotted. A solid black
line indicates the values of $L_a(\lambda)$ and $L_c(\lambda)$ for
optically pure seawater reported by Smith and Baker
\cite{SmithBaker1981}.} \label{fig:Lac_KM4_season}
\end{figure}

Table \ref{tab:KM4_average_ac} gives, for each campaign, the
weighted average values of the absorption and attenuation
coefficients, as a function of wavelength. Weighted average is
calculated from the values of $a$($\lambda$) and $c$($\lambda$),
measured in each deployment at depths between 2850 and 3250 m.
Statistical errors are calculated from the RMS of the observed
distributions. In Figure \ref{fig:Lac_KM4_season} the absorption
and attenuation lengths are shown. During December 1999 and March
2002 campaigns the channel $c$(555) was not properly working; the
same happened to channels $c$(488) during all the campaigns after
May 2002 and to $a$(488) in July 2003. The corresponding data are
not reported here.

\begin{center}
\begin{table}[h]
\caption{Weighted average values of $a$(${\lambda}$) and
$c$(${\lambda}$) measured in \CP~ $KM4$ during different seasons,
in the interval of depth 2850$\div$3250 m.}
\label{tab:KM4_average_ac}
\begin{tabular}{cccccc}
\hline
  coefficient & December 1999& March 2002 & May 2002 & August 2002 & July 2003\\
\hline
  a412 & $0.0145\pm0.0008$ & $0.0151\pm0.0014$ & $0.0187\pm0.0014$ & $0.0205\pm0.0008$ &$0.0127\pm0.0017$\\
  a440 & $0.0164\pm0.0009$ & $0.0166\pm0.0011$ & $0.0160\pm0.0016$ & $0.0148\pm0.0005$ &$0.0126\pm0.0010$\\
  a488 & $0.0210\pm0.0005$ & $0.0212\pm0.0007$ & $0.0189\pm0.0013$ & $0.0181\pm0.0003$ &$ $\\
  a510 & $0.0366\pm0.0007$ & $0.0366\pm0.0007$ & $0.0377\pm0.0013$ & $0.0383\pm0.0005$ &$0.0367\pm0.0008$\\
  a532 & $0.0528\pm0.0004$ & $0.0529\pm0.0006$ & $0.0517\pm0.0010$ & $0.0502\pm0.0005$ &$0.0507\pm0.0006$\\
  a555 & $0.0685\pm0.0006$ & $0.0683\pm0.0007$ & $0.0675\pm0.0008$ & $0.0677\pm0.0005$ &$0.0673\pm0.0005$\\
  a650 & $0.3572\pm0.0009$ & $0.3565\pm0.0010$ & $0.3610\pm0.0004$ & $0.3619\pm0.0004$ &$0.3619\pm0.0003$\\
  a715 & $1.0168\pm0.0003$ & $1.0117\pm0.0014$ & $1.0458\pm0.0003$ & $1.0457\pm0.0002$ &$1.0451\pm0.0003$\\
\hline
  c412 & $0.0319\pm0.0028$ & $0.0331\pm0.0025$ & $0.0351\pm0.0033$ & $0.0327\pm0.0024$ &$0.0334\pm0.0039$\\
  c440 & $0.0287\pm0.0021$ & $0.0302\pm0.0024$ & $0.0281\pm0.0029$ & $0.0283\pm0.0023$ &$0.0288\pm0.0034$\\
  c488 & $0.0315\pm0.0022$ & $0.0329\pm0.0027$ & $               $ & $               $ &$ $\\
  c510 & $0.0406\pm0.0024$ & $0.0414\pm0.0022$ & $0.0436\pm0.0027$ & $0.0450\pm0.0027$ &$0.0459\pm0.0027$\\
  c532 & $0.0497\pm0.0022$ & $0.0510\pm0.0025$ & $0.0577\pm0.0016$ & $0.0584\pm0.0024$ &$0.0574\pm0.0021$\\
  c555 & $               $ & $               $ & $0.0808\pm0.0029$ & $0.0791\pm0.0023$ &$0.0761\pm0.0020$\\
  c650 & $0.3729\pm0.0024$ & $0.3744\pm0.0025$ & $0.3851\pm0.0032$ & $0.3849\pm0.0034$ &$0.3797\pm0.0015$\\
  c676 & $0.4493\pm0.0017$ & $0.4502\pm0.0015$ & $0.4761\pm0.0041$ & $0.4740\pm0.0037$ &$0.4684\pm0.0022$\\
  c715 & $1.0175\pm0.0019$ & $1.0469\pm0.0010$ & $1.0645\pm0.0032$ & $1.0626\pm0.0030$ &$1.0652\pm0.0023$\\
\hline
\end{tabular}
\end{table}
\end{center}

\begin{figure}[h]
\centerline
{\includegraphics[width=9cm]{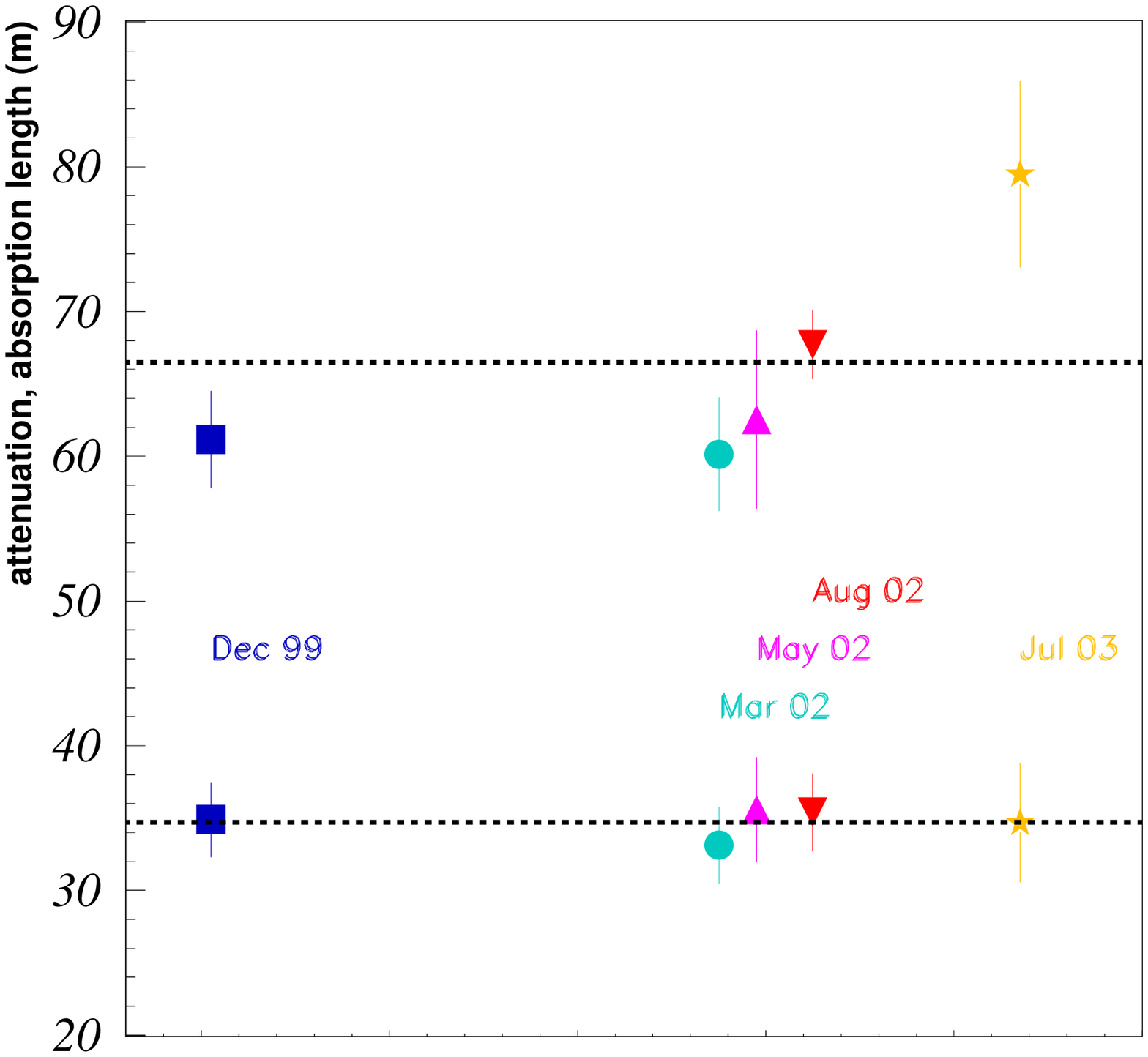}}
\caption{Average attenuation and absorption lengths at
$\lambda$=440 nm measured with the {\it AC9} in $KM4$, at depth
2850$\div$3250 m in December 1999 (blue circle), March 2002 (light
blue square), May 2002 (purple triangle), August 2002 (red
upsidedown triangle) and July 2003 (dark yellow star). The
weighted average values of $L_a$(440) and $L_c$(440) are indicated
by dashed black lines (see text). Statistical errors are shown.}
\label{fig:KM4_season_stability}
\end{figure}

Figure \ref{fig:KM4_season_stability} shows the time dependence of
the average values of $L_a$(440 nm) and $L_c$(440 nm) as a
function of time. The plotted error bars are statistical errors.
The average absorption length, calculated using the values of
Table \ref{tab:KM4_average_ac} weighted with their statistical
errors, is $L_a$ ($\lambda$ = 440 nm) = 66.5 $\pm 8.2_{stat} \pm
6.6_{syst}$ m close to the value of optically pure water. The
weighted average attenuation length is $L_c$($\lambda$ = 440 nm)=
34.7 $\pm 3.3_{stat} \pm 1.8_{syst}$ m close to published values
of ocean waters measured in conditions of collimated beam and
detector geometry \cite{Duntley1963}. The value of $L_c$ measured
in \CP~ is larger than the one reported by Khanaev and Kuleshov
\cite{Khanaev1993} for the NESTOR site. We remind that other
results (by DUMAND \cite{Bradner1981}, NESTOR
\cite{Anassontzis1994} and ANTARES \cite{Aguilar2004}) have been
obtained measuring the \it{effective} light attenuation in
conditions of not collimated geometry, i.e. using a diffused light
source and a large area detector; these results therefore deal
with the \it{effective} attenuation coefficients and cannot be
directly compared with our results.

\section{Conclusions}

The NEMO Collaboration measured, as a function of depth, the
salinity, temperature and inherent optical properties in several
abyssal sites of the central Mediterranean Sea using an
experimental apparatus consisting of an AC9 transmissometer and a
standard CTD probe. In order to compare the water transparency to
\v{C}erenkov light of different sites we have averaged the
measured values of $c(\lambda)$ and $a(\lambda)$ in a range of
about 400 m, at the depths which are suitable for the deployment
of a km$^3$ neutrino telescope. The data of $L_a(\lambda)$
presented for \CP~$KM4$ site are close to the ones reported by
Smith and Baker for optically pure seawater \cite{SmithBaker1981}.
For blue light, the average absorption length is $\simeq$ 67 m,
the average attenuation length is $\simeq$ 35 m. It is worth to
mention that all the measurements reported in this paper have been
carried out over an area of about 10 km$^2$ around the reference
point of $KM4$. We conclude that optical and oceanographic
properties in \CP~$KM4$ site are homogeneous in a large region and
constant over the investigated timescale. The measured absolute
values of IOP and the homogeneity of the water column, for more
than one thousand metres above the seabed, make \CP~ $KM4$ an
optimal site for the installation of the future Mediterranean
km$^3$ underwater neutrino telescope.

\section{Acknowledgements}
This work has been been conducted in collaboration with:
Department of Physical Oceanography INOGS (Trieste), Istituto
Sperimentale Talassografico CNR (Messina) and Istituto di
Oceanografia Fisica CNR (La Spezia). We thank Captains E. Gentile,
V. Lubrano,  A. Patan\'e, the officers and the crew of the R/V
{\it Alliance}, {\it Thetis} and {\it Urania} for their
outstanding experience shown during the sea campaigns.

\end{document}